\documentclass[prl, aps, reprint, superscriptaddress, preprintnumbers, amsfonts, amssymb, amsmath%
, 
]{revtex4-1}
\usepackage{mathrsfs}
\usepackage{graphicx}
\usepackage{hyperref}
\usepackage{slashed}
\hypersetup{
colorlinks, citecolor=[rgb]{0,0,0.8}, linkcolor=[rgb]{0,0,0.8}, urlcolor=[rgb]{0.7,0,0.5}}

\usepackage{tikz}
\usepackage{circuitikz}
\usepackage{ulem}

\newcommand{\be}{\begin{equation}} \newcommand{\ee}{\end{equation}}
\newcommand{\bea}{\begin{equation} \begin{aligned}} \newcommand{\eea}{\end{aligned} \end{equation}}

\newcommand{\cL}{\mathcal{L}}

\newcommand{\cV}{\mathcal{V}}

\DeclareMathOperator{\Tr}{Tr}

\begin{document}

\title{On the Inhomogeneous Phase of the Chiral Gross-Neveu Model}

\date{\today}

\author{Riccardo Ciccone}
\email{rciccone@sissa.it}

\affiliation{SISSA, via Bonomea 265, 34136 Trieste, Italy and \\ INFN, Sezione di Trieste, via Valerio 2, 34127 Trieste, Italy}

\author{Lorenzo Di Pietro}
\email{ldipietro@units.it}

\affiliation{Dipartimento di Fisica, Universit\`a di Trieste, Strada Costiera 11, I-34151 Trieste, Italy \\ INFN, Sezione di Trieste, via Valerio 2, 34127 Trieste, Italy}

\author{Marco Serone}
\email{serone@sissa.it}

\affiliation{SISSA, via Bonomea 265, 34136 Trieste, Italy and \\ INFN, Sezione di Trieste, via Valerio 2, 34127 Trieste, Italy}

\begin{abstract}

There is substantial evidence that the ground state of the 2d chiral Gross-Neveu model, in the presence of a $U(1)$ fermion number chemical potential $\mu$ 
and in the large $N$ limit, is given by a {\it chiral spiral} phase, namely an inhomogeneous phase with a chiral condensate having a spatially periodic phase. We show that the chiral spiral configuration persists at finite $N$  and $T=0$ for any $\mu>0$. Our analysis is based on non-abelian bosonization, that relates the model to a $U(N)_1$ WZW model deformed by current-current interactions. In this description the appearance of the inhomogeneous phase is surprisingly simple.
We also rederive the phase diagram of the large $N$ chiral Gross-Neveu model via a direct diagrammatic computation, finding agreement with previous results in the literature. 

\end{abstract}



\maketitle

\section{Introduction}

Understanding the phase diagram of strongly interacting matter at finite temperature and density is of fundamental importance in physics, 
with applications that range from superconductivity to compact astrophysical objects. In particular, inhomogeneous phases
where the ground state breaks a subset of the translation symmetries can appear in dense systems, for example in the context of ordinary superconductivity \cite{Fulde:1964zz, Larkin:1964wok} and in one-dimensional metals \cite{Gruner:1988zz,Gruner:1994zz}.
Inhomogeneous phases have been conjectured to form in a range of temperatures and matter densities also in relativistic theories such as Quantum Chromodynamics (QCD).
Evidence that cold dense quark matter at large number of colors in QCD might form ``standing chiral waves'', i.e. configurations where only a linear combination of chiral symmetry and translations is linearly realized, has been first provided in \cite{Deryagin:1992rw}. Such chiral waves have been subsequently shown to be disfavoured in actual QCD (three colors)  \cite{Shuster:1999tn}, but the possibility of other inhomogeneous phases in QCD has emerged during the years, see e.g. \cite{Kojo:2009ha, Anglani:2013gfu, Buballa:2014tba} and references therein \footnote{Inhomogeneous phases appear also in holographic models for strongly-coupled gauge theories at finite density, see e.g. \cite{Nakamura:2009tf}.}. 

Due to the limited availability of analytical and numerical tools, it is difficult to rigorously assess the appearance of these phases in QCD.
For this reason, it is interesting to understand the phase diagram of simpler theories that qualitatively resemble QCD. A renowned example is the theory of $N$ massless Dirac fermions in two spacetime dimensions interacting through four-fermion interactions \cite{Gross:1974jv}. These theories, like QCD,  are UV-free and undergo dynamical mass generation in the IR.
One version of the theory, the Gross-Neveu (GN) model, has maximal $O(2N)$ vector symmetry, 
is gapped and shows spontaneous breaking of a ${\bf Z}_2$ chiral symmetry. Another version of the theory, the chiral Gross-Neveu (cGN) model,  
enjoys $U(N)$ vector symmetry and $U(1)$ chiral symmetry. The latter theory is gapless in the IR, having in the spectrum a massless compact boson typical of quasi long-range ordered phases in two dimensions \cite{Berezinsky:1970fr,Berezinsky:1972rfj,Kosterlitz:1973xp}.
The phase diagrams of the GN and the cGN models at large $N$ at finite temperature $T$ and in the presence of a $U(1)$ fermion number chemical potential $\mu$ have first been studied in \cite{Wolff:1985av} and \cite{Barducci:1994cb}, respectively, under the assumption of translational invariance.
This assumption and the corresponding results have later been disputed, and revised large $N$ phase diagrams including inhomogeneous phases have been found \cite{Schon:2000he,Thies:2003kk,Schnetz:2004vr,Basar:2009fg}: a crystal phase in the GN model, with a spatially modulated order parameter at low $T$ and sufficiently large $\mu$, and a phase analogous to the chiral waves, dubbed ``chiral spiral'', in the cGN model, with a spatially periodic phase with $\mu$-dependent period at low $T$ for any $\mu>0$ \footnote{Inhomogeneous phases have also been studied in two-dimensional QCD at finite temperature and chemical potential, see e.g. \cite{Azaria:2016mqb,Lajer:2021kcz}.}. The results of \cite{Thies:2003kk,Schnetz:2004vr,Schon:2000he,Basar:2009fg} are based on solutions of certain self-consistency equations, and apply in the large $N$ limit. 
 In recent years the appearance of inhomogeneous phases at finite $N$ has started to be analyzed using numerical lattice methods (see e.g. \cite{Narayanan:2020uqt,Lenz:2020bxk} for the GN model and \cite{Lenz:2021kzo} for the cGN model) \footnote{See also \cite{Stoll:2021ori, Koenigstein:2021llr} for other numerical studies of the phase diagram of the GN model for finite $N$.}. 

The aim of this letter is to investigate the fate of the phase diagram of the cGN model at finite $N$. Using non-abelian bosonization we show that the chiral spiral phase of the cGN model is present at any finite $N$, at $T=0$ for any $\mu>0$.  The chiral wave corresponds to a quasi long-range ordered gapless phase, in contrast to strict long-range order appearing at large $N$, with a free massless relativistic excitation. In this phase the would-be order parameter is neutral under IR spatial translations that are a linear combination of the original translations and the axial $U(1)$ symmetry; the combination depends on the realization of a certain ${\bf Z}_N$ symmetry in the $SU(N)_1$ Wess-Zumino-Witten (WZW) model deformed by a current-current interaction.

We also rederive  the phases of the cGN model at large $N$ by a diagrammatic computation of the free energy in the presence of the inhomogeneity.

\section{cGN as $J\bar J$ deformation of a WZW model}
The euclidean Lagrangian of the cGN model \cite{Gross:1974jv} is
\begin{equation}\label{eq:LcGN}
    {\cL_{\text{cGN}}=i\psi_{+a}^\dag\partial_-\psi_+^a- i \psi_{-a}^\dag\partial_+\psi_-^a- 
    \frac{\lambda_s}N |\psi_{-a}^\dagger \psi_+^a|^2~, 
    }
\end{equation}
where $\partial_+$ and $\partial_-$ are derivatives with respect to complex coordinates $x^+ =(x^1 - i x^2)/2$, $x^- =(x^1+i x^2)/2$, $a=1,\dots,N$ (summation over repeated indices is understood) and $\psi_\pm^a$ are the two chiral components of a set of $N$ Dirac fermions.
The $O(2N)_+\times O(2N)_-$ global symmetry of the $N$ free Dirac fermions is broken by the interaction to $(U(N)_V\times U(1)_A)/\mathbf{Z}_2$. 
The action of the residual group on the fields is
\be
\begin{aligned}\label{eq:glsymfermions}
    U(N)_V:& \qquad \psi^a_\pm \mapsto \cV^a{}_b\psi^b_\pm ,\qquad \cV\in U(N)~,\\
    U(1)_A: &\qquad  \psi^a_\pm \mapsto {e}^{\pm i \alpha}\psi^a_\pm ,\qquad  \alpha\in{\bf R}~.
\end{aligned}\ee
The scalar-scalar quartic interaction appearing in (\ref{eq:LcGN}) is not the most generic one with the above symmetries. One can also have the vector-vector interaction \footnote{In the vector-like large $N$ limit, the non-trivial scaling with $N$ for a double-trace coupling is $1/N$, rather than $1/N^2$. However scaling this particular coupling as $1/N$ one finds that it has a vanishing $\beta$ function at leading order, and therefore the coefficient of $1/N$ can be consistently set to zero.}
\begin{equation}\label{eq:vvint}
    \cL_{v}= \frac{\lambda_v}{N^2} (\psi_{+a}^\dagger \psi_+^a) (\psi_{-b}^\dagger \psi_-^b)~.
\end{equation}
With a Fierz transformation the Lagrangian (\ref{eq:LcGN}), including the term (\ref{eq:vvint}), can be formulated as
the $U(N)$ generalization \cite{Dashen:1973nhu} of the massless Thirring model \cite{Thirring:1958in}: 
\begin{equation}\begin{aligned}\label{eq:LThirring}
    \hspace{-0.03cm}
    \cL=i \psi_{+a}^\dagger\partial_- \psi_+^a-i \psi_{-a}^\dagger \partial_+ \psi_-^a  +\frac{\lambda J_+^A J_-^A}{N} +\frac{\lambda'J_+ J_-}{N^2}\,,
    \end{aligned}
\end{equation}
where
\begin{equation}
\label{eq:fermicurrents}
  J_\pm^A =\psi^\dagger_{a\pm} (T^A)^a_{~b}\psi_\pm^b~, \quad \quad 
   J_\pm  =\psi^\dagger_{a\pm}\psi_\pm^a~,
\end{equation}
are the $SU(N)$ and $U(1)$ currents, respectively, with $T^A$ the generators of $SU(N)$ in the fundamental representation, normalized as $\Tr(T^AT^B)=\delta^{AB}/2$. The couplings $\lambda_s,\lambda_v,\lambda,\lambda'$ are related by 
\begin{equation}\label{eq:fierzedcouplings}
\lambda=2\lambda_s~,\qquad \lambda'=\lambda_v+\lambda_s~.
\end{equation}

Using non-abelian bosonization \cite{Witten:1983ar,Affleck:1985wb,Affleck:1985wa} (see also \cite{Banks:1975xs}) the theory (\ref{eq:LThirring}) can be equivalently described as a $J\bar{J}$ deformation of a $U(N)_1$ WZW model \footnote{$N$ free Dirac fermions can actually be bosonized keeping manifest the full $O(2N)_+\times O(2N)_-$ symmetry of the theory. The $U(N)$ prescription \cite{Knizhnik:1984nr} is more convenient here, since the $J\bar J$ deformation preserves only the unitary symmetry.}. Locally the degrees of freedom are a $SU(N)$ matrix $U$ and a scalar $\phi$ of radius $R =\sqrt{N}$ \footnote{In our conventions, the self-$T$-dual radius for the compact boson is $R=\sqrt{2}$.} parametrizing the $U(1)$ factor, in terms of which the Lagrangian is
\begin{equation}\label{eq:bosonizedthirring}
    {\cal L}[U,\phi]={\cal L}_0[U,\phi] + \frac{\lambda  J^A_+J^A_-}N + \frac{\lambda'  J_+J_-}{N^2}~,
\end{equation}
where
\begin{align}
\begin{split}
{\cal L}_0[U,\phi]  & =  \frac{1}{8\pi} \partial_+ \phi \,\partial_- \phi \\ & \hspace{-0.6cm} + \frac1{8\pi} {\rm Tr} (\partial_+ U^\dag \partial_- U+\partial_- U^\dag \partial_+ U)+ {\cal L}_{\text{WZ}}^{SU(N)_1}~,
\end{split}
\end{align}
is the Lagrangian of the undeformed WZW model, with ${\cal L}_{\text{WZ}}^{SU(N)_1}$ being the level $k=1$ $SU(N)$ Wess-Zumino term, which is the boundary term of the three-dimensional topological Lagrangian
\begin{equation}
   \frac1{12\pi}\epsilon^{ijk}\Tr(U^\dag\partial_iUU^\dag\partial_jUU^\dag\partial_kU),
\end{equation} and 
\begin{align}
    J^A_+  =\frac{i}{2\pi}\Tr\left( U^\dag \right.&\left. \!\!\!(\partial_+U)T^A\right),~ J^A_-=\frac{i}{2\pi}\Tr\left((\partial_-U) U^\dag T^A\right),\nonumber\\
  &J_\pm  =- \frac{\sqrt{N}}{4\pi}\partial_\pm\phi~,
    \label{eq:bosecurrents}
\end{align}
are the bosonized $SU(N)$ and $U(1)$ currents. 

The $J^A_+J^A_-$ deformation breaks the $(SU(N)_+\times SU(N)_-)/{\bf Z}^V_N$ symmetry of the $SU(N)_1$ WZW model,
\begin{equation}
    \!{SU(N)_+\times SU(N)_-}:\,  U\mapsto g_-Ug_+^\dag,\, g_\pm\in SU(N)\,,
\end{equation}
to $(SU(N)_V/{\bf Z}^V_N)\times {\bf Z}^A_N$. Here ${\bf Z}^V_N$ denotes a transformation in the center of the diagonal group $SU(N)_V$ which leaves $U$ invariant, while ${\bf Z}^A_N$ denotes a transformation in the center of only one of the two $\pm$ factors (it does not matter which one) and acts as a phase on $U$. Explicitly,
\begin{equation}\begin{aligned}
    SU(N)_V/\mathbf{Z}_N^V&:\quad  U\mapsto VUV^\dag,\quad V\in SU(N)~,\\
    \mathbf{Z}_N^A&:\quad U\mapsto e^{2\pi i k/N}U,\quad k\in\mathbf{Z}_N~.
\end{aligned}
\end{equation}
In the compact boson sector instead, the operator $J_+J_-$ is not an interaction, as
\begin{equation} \label{eq:scalarJJ}
    \frac{\lambda'J_+J_-}{N^2}=\frac{\lambda'}{16\pi^2N}\partial_+ \phi\partial_-\phi~.
\end{equation}
The compact scalar $\phi$ and the $SU(N)$ matrix $U$ are completely decoupled in the Lagrangian \eqref{eq:bosonizedthirring}. However, since $U(N)\simeq [SU(N)\times U(1)]/{\bf Z}_N $, to obtain the deformed $U(N)_1$ model from \eqref{eq:bosonizedthirring} we further need to gauge the diagonal ${\bf Z}_N$ symmetry between ${\bf Z}^A_N$ and a ${\bf Z}^\phi_N$ subgroup of the shift symmetry of the compact boson. A similar subtlety arises in the identification of the deformed $U(N)_1$ model with the original fermionic theory \eqref{eq:LThirring}. Since the fermionic theory depends on the spin structures while the bosonic one does not, to obtain a full equivalence one needs to gauge a ${\bf Z}_2$ symmetry (see e.g. \cite{Bhardwaj:2016clt, Karch:2019lnn, Ji:2019ugf}). 
The gauging of these discrete symmetries will not affect the calculation of the free energy on ${\bf R}^2$ --where we do not have twisted sectors-- as a function of $\mu$. However both the gauging of ${\bf Z}_N$ and of ${\bf Z}_2$ can play a role in the study of the thermal partition function, because of the non-trivial topology of the background. It would be interesting to study these issues in more detail. 

The value of $\lambda'$ is an RG invariant, because it only changes the radius of the compact boson. The coupling $\lambda$, instead, runs and becomes strong in the IR. Equivalently, the couplings $\lambda_s,\lambda_v$ must both run, with $\beta_{\lambda_v}=-\beta_{\lambda_s}$ \cite{Bondi:1989nq}. The coupling $\lambda_v$ at finite $N$ has then to be included, as it is anyway generated radiatively.

The impossibility of a trivially gapped theory could be anticipated by the presence of a 't Hooft anomaly between the $U(1)_V$ and $U(1)_A$ currents. In the bosonized version of the theory, this 't Hooft anomaly is reproduced by the compact scalar, where $U(1)_A$ and $U(1)_V$ act as a shift on $\phi$ and on its dual, respectively.

\section{Chiral spiral at finite $N$}

Let us introduce a chemical potential for the $U(1)_V$ charge by adding to the Lagrangian (\ref{eq:LThirring}) the term 
\begin{equation}\label{eq:Lmu}
    \cL_\mu= \mu(\psi^\dag_{+a}\psi_+^a+\psi^\dag_{-a}\psi_-^a)~.
\end{equation}
Upon bosonization, this maps simply to 
\begin{equation}\label{eq:Lmubose}
    \cL_\mu=-\mu\frac{\sqrt{N}}{2\pi}\partial_1\phi~.
\end{equation}
The term \eqref{eq:Lmubose}, which  does not depend on the $SU(N)$ degrees of freedom, provides an expectation value for $\partial_1\phi$. By adding (\ref{eq:Lmubose}) to the Lagrangian (\ref{eq:bosonizedthirring}) we get that the effective action for $\phi$ is minimized on configurations with
\begin{equation} \label{eq:dephivev}
    \langle \partial_1 \phi\rangle=2\mu\sqrt{N}\left(1+\frac{\lambda'}{2\pi N}\right)^{-1} \equiv 2 \mu' \sqrt{N}~. 
\end{equation}
The difference of the zero-temperature free energy density per flavour between the configuration with and without the expectation value for $\partial_1\phi$ is
\begin{equation}\label{eq:freeenergyshift}
    \delta F = 
    -\frac{\mu^2}{2\pi}\left(1+\frac{\lambda'}{2\pi N}\right)^{-1}~,
\end{equation}
showing that the former is favored on ${\bf R}^2$. 

Using the bosonization identity  $\psi^\dag_{-a}\psi_+^b= U_a{}^b e^{{i\phi/\sqrt{N}}}$ \cite{Witten:1983ar,Affleck:1985wa,Affleck:1985wb} (omitting a scheme-dependent renormalization mass scale),
the two-point function of the fermion bilinear
in terms of the $SU(N)$ and free scalar degrees of freedom is
    \begin{align}\label{eq:twopoint}
    &\langle \psi_{-a}^\dag\psi_+^a(x)\psi_{+b}^\dag\psi_-^b(0)\rangle=\\
    \nonumber&\quad\langle\Tr U(x) \Tr U^\dag(0)\rangle e^{2i\mu'x^1}\langle e^{i\frac{\delta\phi(x)}{\sqrt{N}}}e^{-i\frac{\delta\phi(0)}{\sqrt{N}}}\rangle\,,
    \end{align}
        where $\delta\phi$ denotes excitations of $\phi$ around (\ref{eq:dephivev}) and we have used the decoupling of the two sectors to factorize the correlator.  
        
        Let us now assume that the operator $\Tr U$ has a non-vanishing expectation value (at zero temperature) in the $SU(N)_1$ theory deformed by the current-current interaction.  Then, in the limit $|x|\to\infty$, (\ref{eq:twopoint}) approaches
        \begin{equation}
            |\langle\Tr U\rangle|^2 e^{2i\mu'x^1}|x|^{-\frac{2}{N(1+\lambda'/2\pi N)}}\,,
        \end{equation}
        that is, it decays with power-like behavior times an oscillating factor. The latter is consequence of the chiral spiral configuration, whereas the former is the hallmark of quasi long-range order. Only a diagonal subgroup of the $U(1)_A$ symmetry and spatial translations preserves the would-be order parameter $\Tr U e^{2i\mu' x^1}$.
Recalling that the $\psi_{-a}^\dag\psi_+^a$ bilinear carries $U(1)_A$ charge $+2$, this subgroup is a $U(1)_A$ transformation of parameter $\alpha$ accompanied by a translation with parameter $\delta x^1 = -\alpha/\mu'$. The condensation of $\Tr U$ breaks completely the global center symmetry $\mathbf{Z}^A_N$ of the $SU(N)_1$ theory. 
In the next section we will see that indeed in the large $N$ limit the fermion bilinear gets a chiral spiral condensate (in the strict sense), giving evidence in favor of this assumption.
If the $\mathbf{Z}^A_N$ symmetry is unbroken in the $SU(N)_1$ theory ($\langle\Tr U\rangle=0$), an operator of the cGN theory with a spatially modulated expectation value must contain only vertex operators of the form $e^{i \,k N \, \phi/\sqrt{N}}$, with $k\in\mathbf{Z}$. This ensures that the operator is invariant under $\mathbf{Z}^\phi_N$, without the need of a compensating factor charged under $\mathbf{Z}^A_N$ from the $SU(N)_1$ sector (recall that the diagonal subgroup of these two $\mathbf{Z}_N$ is gauged, so any physical operator must be neutral under it). For instance, the quasi-long range order could be detected in the two-point function of the $N$-th power of the fermion bilinear $(\psi_{-a}^\dag\psi_+^b)^N$. In this case the 
 $U(1)_A$ transformation of parameter $\alpha$ needs to be accompanied by a translation with parameter $\delta x^1 = - N\alpha/\mu'$. There are also intermediate possibilities in which only a nontrivial subgroup $\mathbf{Z}^A_{N'} \subset \mathbf{Z}^A_{N}$ is preserved, when $N$ is a multiple of $N'$.

The vacuum is then in a so-called ``chiral spiral" configuration, where only a linear combination of the $U(1)_A$ symmetry and of spatial translations preserves the would-be order parameter. Such combination depends on the realization of the $\mathbf{Z}^A_N$ symmetry in the vacuum of the deformed $SU(N)_1$ theory.  Excitations on top of the chiral spiral are gapless and have a relativistic dispersion relation. 

These results apply in ${\bf R}^2$, namely at $T=0$ and at infinite spatial length. It is reasonable to expect that some chiral spiral configuration persists for $T>0$ for sufficiently small temperatures in a thermal circle, when fermions are taken anti-periodic along the time direction. Discrete symmetries are however always restored in two dimensions at $T > 0$.
As a result, $\langle \Tr U\rangle$ vanishes and the chiral spiral cannot be detected by looking at the fermion bilinear but possibly only at  $\mathbf{Z}_N^A$-invariant order parameters. Similarly, on a spatial circle of length $L$ chiral spirals can possibly occur only for quantized values of the step.
It would be interesting to establish more firmly the fate of the chiral spiral configuration for both $T>0$ and finite spatial length $L$ at finite $N$. 

\section{Chiral spiral at large $N$}

The cGN model in the large $N$ limit is conveniently studied by introducing a complex Hubbard-Stratonovich (HS) field $\Delta$ rather than reformulating the theory as a $J\bar J$ deformation of a WZW model. At large $N$ the  $\mathbf{Z}^A_N$ symmetry can be spontaneously broken at finite $T$, because the usual no-go theorems do not apply in this limit \footnote{For the same reason, $\mathbf{Z}^A_N$ can be broken even if it looks like a continuous $U(1)$ symmetry at infinite $N$}. In this section we derive the critical temperature $T_c$ of the cGN model, reproducing the result of \cite{Thies:2003kk,Schnetz:2004vr,Schon:2000he,Basar:2009fg} with a different method. To this end, we compute the free energy density per flavour (which for simplicity is called free energy in what follows) both for a homogeneous condensate and for a inhomogeneous one, assumed to have a chiral spiral form analogous to the one found above at finite $N$ and $T=0$. We show that at low temperatures $T<T_c$ the  latter minimizes the free energy also at large $N$; for $T>T_c$ the symmetry-preserving configuration, in which fermions are massless, is recovered.

We take the 't Hooft limit $N\to\infty$ with $\lambda_s$, $\lambda_v$ fixed. The cGN model at large $N$ is recovered for $\lambda_v=0$.
The free energy of the cGN model at large $N$ is given by
\begin{equation}
    F=
    \frac{\overline{|\Delta|^2}}{\lambda_s}-\Tr\log\left(\slashed{\partial}+  \Delta P_+ + \Delta^*P_-\right)~,
\end{equation}
where $\overline{|\Delta|^2}$ denotes the spacetime average of the square modulus of the condensate, and $P_\pm=(1\pm \gamma_*)/2$ denote chiral projectors, $\gamma_*$ being the 2d chirality matrix. 

Under a $U(1)_A$ transformation the HS field $\Delta=\rho \, e^{i\theta}$ transforms as $\theta\mapsto \theta +2\alpha$, $\alpha\in{\bf R}$.
At large $N$ we can then identify 
    $\theta= {\phi}/{\sqrt{N}}$,
where $\phi$ is the compact scalar in the bosonization of the model.
In terms of the HS field $\Delta$, the chiral spiral configuration \eqref{eq:dephivev}  reads 
$\Delta=Me^{2iqx}$,
where $q=\mu$ at large $N$ (at fixed $\lambda_s$ and $\lambda_v = 0$), and we denote $x= x^1$. However we keep $q$ generic, and compute the free energy in these configurations.

We can perform a perturbative expansion in the coupling $\lambda_s$, or equivalently in $M$. Neglecting irrelevant constant terms, we have 

\begin{align}
    \nonumber F(M,q)&= 
     \frac{M^2}{\lambda_s}  +\sum_{n=1}^\infty \frac{\mathrm{Tr}[-\slashed{\partial}^{-1}M(e^{2iqx}P_++e^{-2iqx}P_-)]^n}n\\
    &=\frac{M^2}{\lambda_s}  -\sum_{m=1}^\infty \frac1{2m}\times 2\tikz[baseline=.05ex]{
    \node[circle,minimum size=30pt,draw](c) at (0,0){};
    \draw (c.west)--(c.west) node[currarrow,sloped,midway] {};
    \draw (c.north west) node[rotate=0] {\color{red} $\oplus$};
    \draw (c.north west) node[anchor=south east](t) {$1_+$};
    \draw (c.north)--(c.north) node[currarrow,midway] {};
    \draw (c.north east) node[rotate=0] {\color{blue} $\ominus$};
    \draw (c.north east) node[anchor=south west] {$1_-$};
     \draw (c.east)--(c.east) node[currarrow,sloped,midway,rotate=180] {};
     \draw (c.south west) node[rotate=0] {\color{blue} $\ominus$};
    \draw (c.south west) node[anchor=north east] {$m_-$};
     \draw (c.south east) node[anchor=north, rotate=30](b) {$...$};
     }~.
    \end{align}
We can interpret the traces diagrammatically as follows. Oriented lines denote massless fermion propagators. Each insertion of $P_\pm$, denoted by ${\color{red}\oplus},{\color{blue}\ominus}$ respectively, brings an insertion of mass $M$ and spatial momentum $\mp2q$. By momentum conservation in the loop there has to be equal number of $P_\pm$ insertions, which must be alternated because $P_\pm\gamma^\mu P_\pm=0$, therefore only terms with even $n=2m$ contribute. Invariance under ${\color{red}\oplus}\leftrightarrow{\color{blue}\ominus}$ exchange gives the extra factor of $2$.

Letting $p_{1\mp}=p_1\mp2q$, at $T=\mu=0$ one has
\begin{align}
    &\sum_{m=1}^\infty\frac{1}{m}
    \tikz[baseline=.05ex]{
    \node[circle,minimum size=30pt,draw](c) at (0,0){};
    \draw (c.west)--(c.west) node[currarrow,sloped,midway] {};
    \draw (c.north west) node[rotate=0] {\color{red} $\oplus$};
    \draw (c.north west) node[anchor=south east](t) {$1_+$};
    \draw (c.north)--(c.north) node[currarrow,midway] {};
    \draw (c.north east) node[rotate=0] {\color{blue} $\ominus$};
    \draw (c.north east) node[anchor=south west] {$1_-$};
     \draw (c.east)--(c.east) node[currarrow,sloped,midway,rotate=180] {};
     \draw (c.south west) node[rotate=0] {\color{blue} $\ominus$};
    \draw (c.south west) node[anchor=north east] {$m_-$};
     \draw (c.south east) node[anchor=north, rotate=30](b) {$...$};
     } =\qquad\qquad \\
     &=\int\frac{d^2p}{(2\pi)^2}\log\left(1+\frac{M^2(p_1^2+p_2^2-q^2+2iqp_2)}{(p_{1-}^2+p_2^2)(p_{1+}^2+p_2^2)}\right)~.\nonumber
\end{align}
At $T>0$ we need to replace $\int \frac{dp_2}{2\pi}\to T\sum_{p_2}$, with $p_2=\pi(2n+1)T$, and at $\mu\neq0$,  $p_2\to p_2+i\mu$. After $p_1$ integration we have
\begin{equation}
        \!\!\!F(M,q)=
        \frac{M^2}{\lambda_s} -T\sum_{p_2}\left[\sqrt{M^2+p_{2}^2(q)}-\sqrt{p_{2}^2(q)} \right]~,
\end{equation}
where $p_{2}(q) = \pi (2n+1) T + i (\mu -q)$.
This expression contains divergences and needs to be regularized.  We can use the $T=\mu=0$ gap equation
\begin{equation}
    \frac1{\lambda_s}=\int_{-\Lambda}^\Lambda\frac{d p_2}{2\pi}\int\frac{d p_1}{2\pi}\frac1{p^2+M_0^2}\approx \frac1{2\pi}\log\left(\frac{2\Lambda}{M_0}\right)~,
\end{equation}
to trade the dependence on the bare coupling $\lambda_s$ for the UV cutoff $\Lambda$. Here $M_0$ is the vacuum expectation value of $|\Delta|$ at $T=\mu=0$. Similarly, we regularize the sum by putting a cutoff $p_{2,{\rm max}}=2\pi n_{{\rm max}}T$ over the Matsubara frequencies and relate it to the UV cutoff by letting $\Lambda=p_{2,{\rm max}}$. The resulting expression is now finite in the limit $n_{{\rm max}}\to\infty$, and we obtain
\begin{align}
    &F(M,q)= 
    \frac{M^2}{2\pi}\left[\log\left(\frac{4\pi T}{M_0}\right)+{\rm Re}\,\psi\left(\frac12 + i\frac{\mu-q}{2\pi T}\right)\right]\nonumber\\
    &\;\;-T\sum_{p_2}\left[\sqrt{p^2_{2}(q)+M^2}-\sqrt{p^2_{2}(q)}-\frac{M^2}{2\sqrt{p^2_{2}(q)}}\right]~,
\end{align}
where $\psi$ is the digamma function.

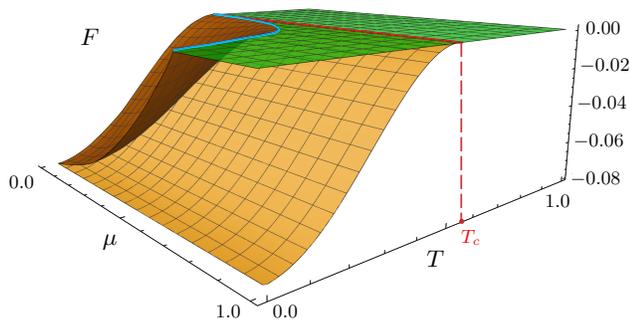
\begin{figure}[t]
    \centering
    \begin{tikzpicture}
    \node[anchor=south west,inner sep=0] (image) at (0,0) {\includegraphics[width=0.45\textwidth]{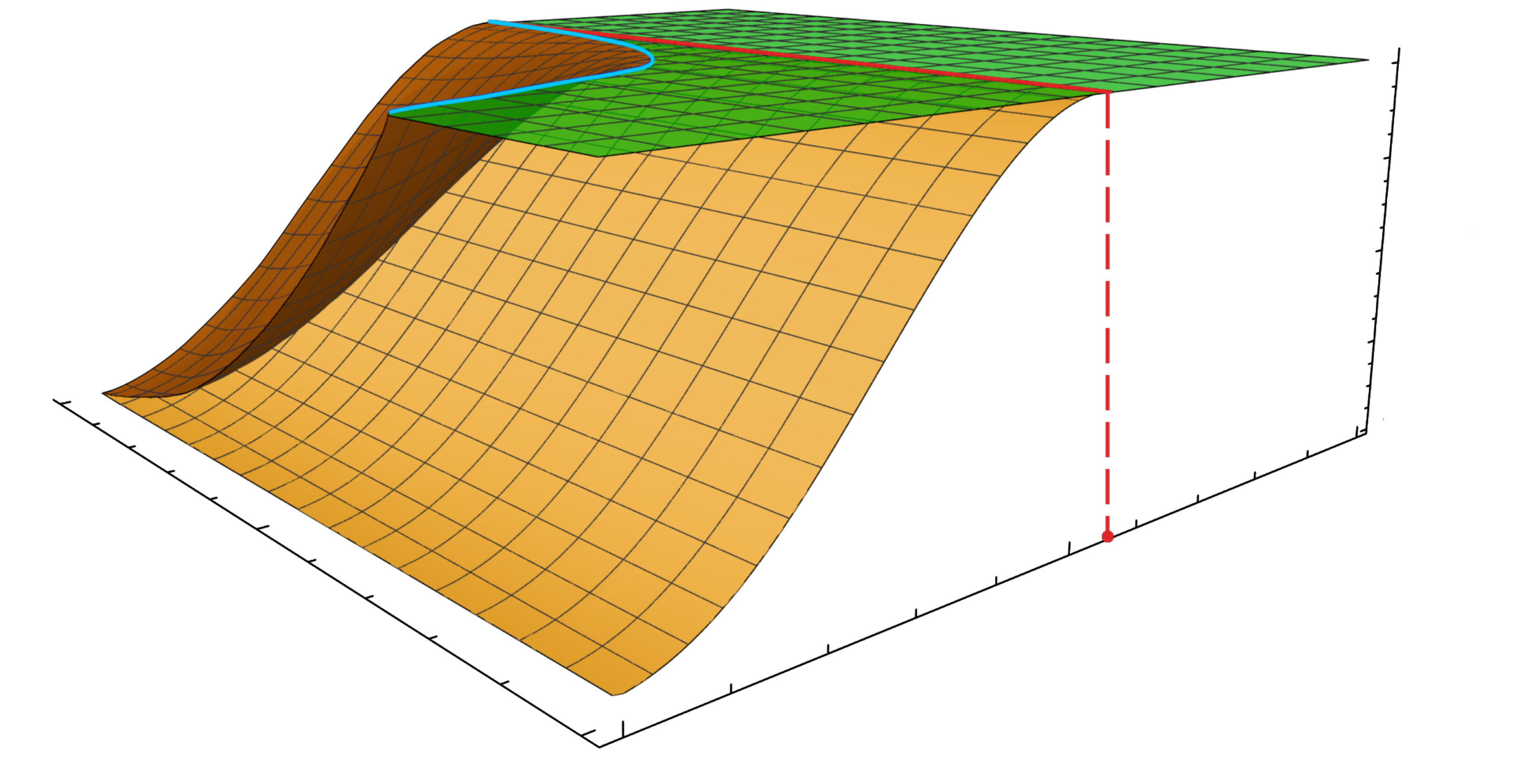}};
    \begin{scope}[x={(image.south east)},y={(image.north west)}]
        \node [anchor=north east,scale=0.8] at (0.04,0.48) {$0.0$};
        \node [anchor=north east,scale=0.8] at (0.38,0.07) {$1.0$};
        \node [anchor=north west,scale=0.8] at (0.41,0.07) {$0.0$};
        \node [anchor=north west,scale=0.8,text=red] at (0.72,0.31) {$T_c$};
        \node [anchor=north,scale=0.8] at (0.89,0.42) {$1.0$};
        \node [anchor=west,scale=0.8] at (0.9,0.45) {$-0.08$};
        \node [anchor=west,scale=0.8] at (0.905,0.57) {$-0.06$};
        \node [anchor=west,scale=0.8] at (0.91,0.69) {$-0.04$};
        \node [anchor=west,scale=0.8] at (0.915,0.81) {$-0.02$};
        \node [anchor=west,scale=0.8] at (0.925,0.93) {$0.00$};
        \node [anchor=north east] at (0.18,0.3) {$\mu$};
        \node [anchor=north west] at (0.66,0.25) {$T$};
        \node [anchor=east] at (0.15,.9) {$F$};
    \end{scope}
    \end{tikzpicture}
    \caption{The large $N$ free energy $F$ as a function of  $\mu$ and $T$, in units of $M_0$. At low $T$, the chiral spiral configuration (yellow) is favored with respect to the homogeneous configuration (brown) and the chirally symmetric phase (green). At $T=T_c$, there is a second order phase transition (red line) dividing the chiral spiral phase from the chirally symmetric phase. Assuming homogeneity one finds a different phase transition line (blue line).}
    \label{fig:freeenergycomparison}
\end{figure}

We compare the minimum of the free energy over the whole parameter space,
    $F_{{\rm cs}}=\min_{M,q}F(M,q)$,
with the one computed assuming translational invariance,
    $F_{{\rm hom}}=\min_{M}F(M,0)$.
For any given value of $M$, minimization with respect to $q$ of the former is achieved at
    $q=\mu$,
in agreement with \eqref{eq:dephivev}. The $\mu$-dependence drops completely in $F_{\text{cs}}$, and 
$F_{\text{cs}}=F_{\text{hom}}(\mu=0)$.
The residual minimizations over $M$ are performed numerically and the result is plotted as a function of $(\mu,T)$ in Figure \ref{fig:freeenergycomparison}.
The chiral spiral configuration $\Delta=M(T)e^{2i\mu x}$ is always favored with respect to the homogeneous configuration $\Delta=M(\mu,T)$; 
moreover, one has $M(T)=0$, i.e. the symmetric massless phase is recovered, for
$T\geq T_c=M_0{e^\gamma}/{\pi}$, 
where a second order phase transition occurs \footnote{The value of $T_c$ is renormalization scheme-dependent.}.
Inhomogeneous vacuum configurations of this kind were already found in \cite{Thies:2003kk,Schnetz:2004vr,Schon:2000he,Basar:2009fg} at large $N$ by using a Hartree-Fock approach and by solving the gap equation for  inhomogeneous condensates. Here we provided a streamlined computation of the free energy based on a resummation of Feynman diagrams, and our findings are in agreement with these previous works.

More recently, inohomogeneous phases of the cGN model at finite $N$ have been found using lattice simulations (also for small but non-vanishing $T$).
Though it is not clear to us to what extent these lattice results persist when the continuum and infinite volume limits are taken, it is worth noticing that
the different slopes of the two linear fits in Fig. 8 of \cite{Lenz:2021kzo} ($\langle \partial_1 \phi \rangle /\sqrt{N}$ as a function of $\mu$  in our notation) is in qualitative agreement with our analytic finding \eqref{eq:dephivev}, which predicts that the linear slope decreases as $N$ decreases.

Summarizing, using non-abelian bosonization, we have found that inohomogeneous phases persist for any finite $N$ and any $\mu$ in the cGN model at $T=0$. 
We have also presented a simple derivation of the phase diagram of the large $N$ cGN model as a function of $\mu$ and $T$ and reproduced the findings of \cite{Thies:2003kk,Schnetz:2004vr,Schon:2000he,Basar:2009fg}.

\vskip 5pt

\begin{acknowledgements}
We thank Francesco Benini, Christian Copetti and Alberto Nicolis for discussions. Work partially supported by INFN Iniziativa Specifica ST\&FI. LD also acknowledges support by the program ``Rita Levi Montalcini'' for young researchers. 
\end{acknowledgements}

\bibliography{Refs}
\end{document}